\documentclass{iauc}
\usepackage{graphicx}
\pubyear{2005}
\volume{199}
\pagerange{1--}
\jname{Probing Galaxies through Quasar Absorption Lines}
\editors{P. R. Williams, C. Shu, and B. M\'{e}nard, eds.}

\title[Obscuration of DLAs] 
{Dust obscuration of DLA galaxies}

\author[Giovanni Vladilo]   
{Giovanni Vladilo$^1$}%

\affiliation{$^1$Istituto Nazionale di Astrofisica,
Osservatorio Astronomico di Trieste, Trieste, Italy \break 
email: vladilo@ts.astro.it\\[\affilskip]
}

\begin{document}

\maketitle

\begin{abstract}
We show that the extinction of quasar absorbers
  increases
exponentially with the logarithmic column density of any volatile metal
(e.g. zinc), 
with a charateristic turning point above which 
the   quasars are severely dimmed. 
We derive the relation between extinction, HI column density ($N_\mathrm{HI}$), 
metallicity ($Z \equiv N_\mathrm{ZnII}/N_\mathrm{HI}$) and
fraction of iron in dust [$f_\mathrm{Fe}(Z)$]
in  Damped Lyman $\alpha$ (DLA) systems. 
We use this relation to estimate the effect of dust obscuration
on the statistical distributions of $N_\mathrm{HI}$ and $Z$ 
measured in magnitude-limited surveys of DLAs.
In the redshift range where the measurements
of zinc column densities have sufficient statistics ($1.8 \leq z \leq 3$)
we find that the  obscuration bias affects the shapes of the observed distributions.
The metallicity distribution is particularly affected by the bias,
which hampers the detection of DLA galaxies with near solar metallicity.
Our results perfectly reproduce,  without tuning  the dust parameters,   
the DLA
observational threshold $\log N(\mathrm{ZnII}) \, \mathrm{[cm^{-2}]}
\lesssim 13.15$ 
found by Boiss\'e and collaborators in 1998, in terms of a rapid rise of the
obscuration.
Our predictions of the effects of  the   bias on the
statistics of DLAs are consistent with observational results 
obtained from  unbiased surveys of radio-selected quasars.   
%
%
%
%

\end{abstract}

\firstsection 
\section{Introduction}

A variety of astronomical observations indicate that
interstellar dust is a pervasive galactic component, not only in the Milky Way and
nearby galaxies, but also in the high-redshift Universe (Meurer 2004). 
%
Among quasar absorbers,
Damped Lyman $\alpha$ (DLA) systems are the best candidates to host dust
because of  
 their high HI column density, which  
shields the grain from ionizing photons, and of their     metal content,
higher than in other absorbers.
If dust is present in DLA systems, it will scatter and absorb the radiation
of the background quasar, dimming its apparent magnitude (extinction)
and changing the slope of its energy distribution (reddening). 
In the most extreme cases, the extinction may obscure the quasar
leading to a selection effect, the obscuration bias, which may prevent the
detection of metal-rich, dusty DLAs (Ostriker \& Heisler 1984, Fall \& Pei 1993).

Measuring    the
extinction or reddening of DLAs
is a difficult  task
owing to the uncertain, variable continuum of the quasars.
%
Detection of quasar reddening from intervening DLAs was
first claimed by Pei et al. (1991).  
Recent investigations based on large data sets find     
reddening from   metal absorbers at redshift $1 \leq z \leq 1.9$ (Khare et al. 2005), 
but   not  
from  DLAs   at $z \sim 2.8$ (Murphy \& Liske 2004).  
Differential reddening in two lines of sight of
 one   DLA at $z =0.9$ has been reported 
 from the study of a gravitationally lensed quasar (Lopez 2005).  
Evidence of the obscuration bias has been searched by comparing
the statistics of  magnitude-limited  surveys
with those of complete surveys of DLAs lying in front of radio-selected quasars. 
The CORALS survey has provided a marginal signal
of obscuration
in the absorber number density, the neutral gas content 
(Ellison et al. 2001) and the metallicity (Ellison 2005), but only
at $\approx 1 \sigma$ level in each case.

The lack of DLA systems with high values of metallicity and HI
column density, above the   threshold 
$\log N(\mathrm{ZnII})$ [cm$^{-2}$] $\simeq 13.15$,
was tentatively attributed to the effect of dust obscuration
(Boiss\'e et al. 1998).
The original motivation of the present work was to understand if
that   threshold 
can be quantitatively explained in terms of a fast rise of the DLA extinction, 
in which case it  would represent an important clue of the obscuration effect.
With this goal in mind, we first derived the relation between
interstellar extinction and $N(\mathrm{ZnII})$.
We then used this relation to quantify the effect of the bias on
the   distributions of column densities and metallicities  of DLAs. 
Finally, we applied our method
to the current sample DLAs with ZnII measurements.
These three parts of our work are briefly outlined
in the next three sections. 
More details   can be found in the full  
    paper (Vladilo \& P\'eroux 2005). 

%
%

\section{The extinction law}

The extinction in a photometric band
with effective wavelength $\lambda$, $A_\lambda$, is  
related to the line-of-sight
zinc column density, 
$N_\mathrm{Zn} \equiv N(\mathrm{ZnII})$, by
means of the   relation
\begin{equation}
A_\lambda = a_\lambda \, 10^{  \, \log N_\mathrm{Zn} } ~
\label{A_Zn}
\end{equation}
where
\begin{equation}
a_\lambda =   
1.817 \times 10^{-24} ~ \mathrm{A_{Fe}} \times g_\lambda \times
 { f_\mathrm{Fe} \over (1-f_\mathrm{Zn})}  \times
 \left( { {\mathrm{Fe} \over \mathrm{Zn} } } \right)  ~.
 \label{a_lambda}
\end{equation}
These relations can be derived by comparing the
column densities of a refractory and a volatile element in
a given line of sight (we adopted
iron and zinc, respectively).
The expression (\ref{a_lambda}) is composed of:
(1)  constant factors
(e.g. A$_\mathrm{Fe}$ is the atomic mass of iron);
(2) a   dust grains parameter, $g_\lambda$;
(3) the fraction   of iron and zinc atoms which are
incorporated in the dust,
$f_\mathrm{Fe}$ and $f_\mathrm{Zn}$;  
(4) the relative abundance
of iron and zinc in the medium, (Fe/Zn).
This last term   is expected to be constant,
close to the solar value, in the local ISM.
The term 
$ { f_\mathrm{Fe} / (1-f_\mathrm{Zn})}$ is approximately constant
and $\approx 1$ by definition of  refractory and volatile elements.
The only factor expected to vary  is       
\begin{equation}
g_\lambda = s ~
{ Q_{e,\lambda}  \over 
r_\mathrm{gr}  \, \varrho_\mathrm{gr} \,   
X^\mathrm{d}_\mathrm{Fe}  }  ~,
\end{equation}
where $s$ is a geometrical factor ($=3/4$ for spherical grains),
$Q_{e,\lambda}$ the extinction efficiency factor (Spitzer 1978),
$r_\mathrm{gr}$ the grain radius, $\varrho_\mathrm{gr}$ the grain
density, and $X^\mathrm{d}_\mathrm{Fe}$ the abundance by mass
of iron in the dust. 

\begin{figure} 
\hskip 0.7cm
 \includegraphics[width=5.2cm,angle=0]{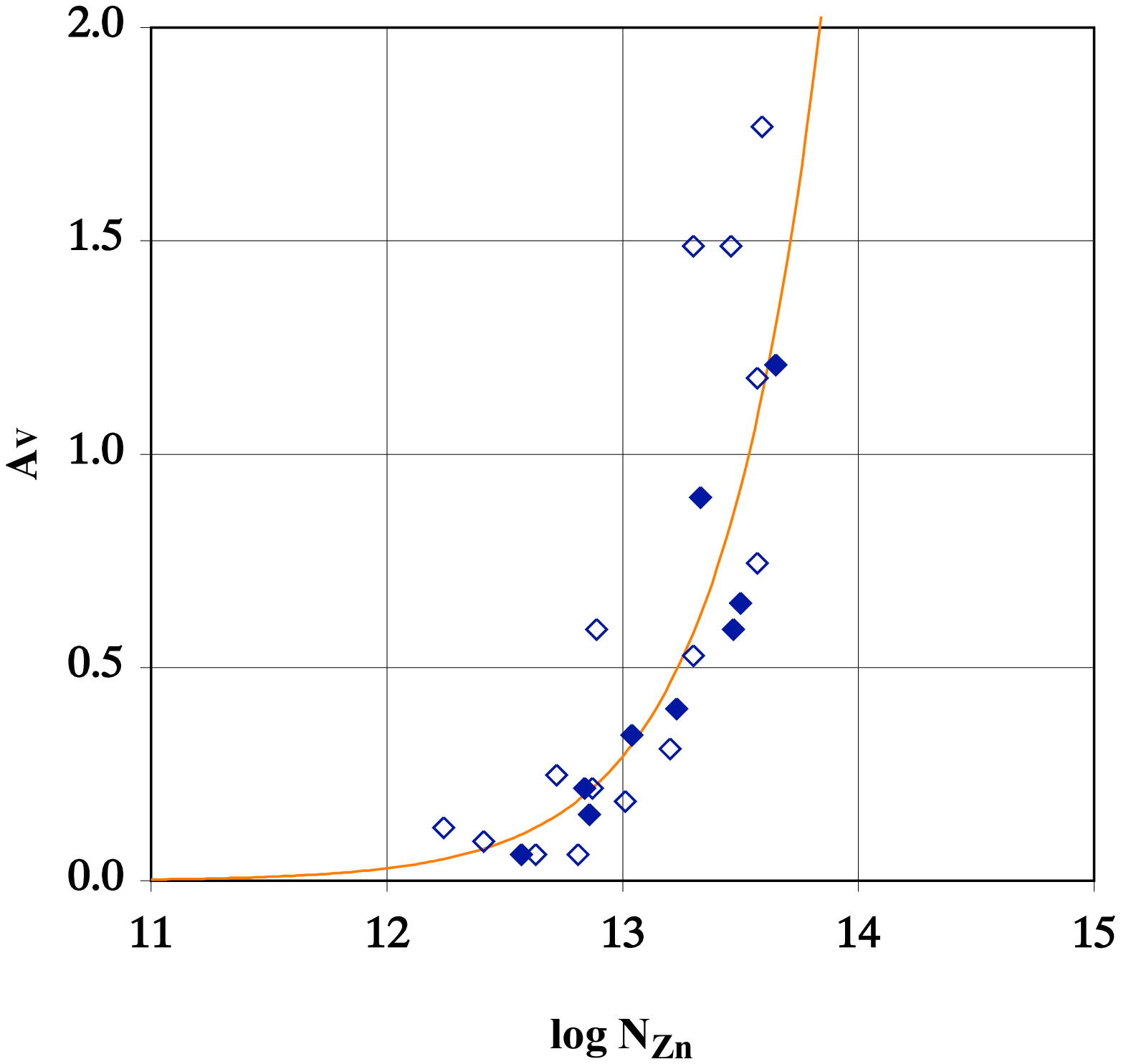}
\hskip 0.7cm
\includegraphics[width=5.4cm,angle=0]{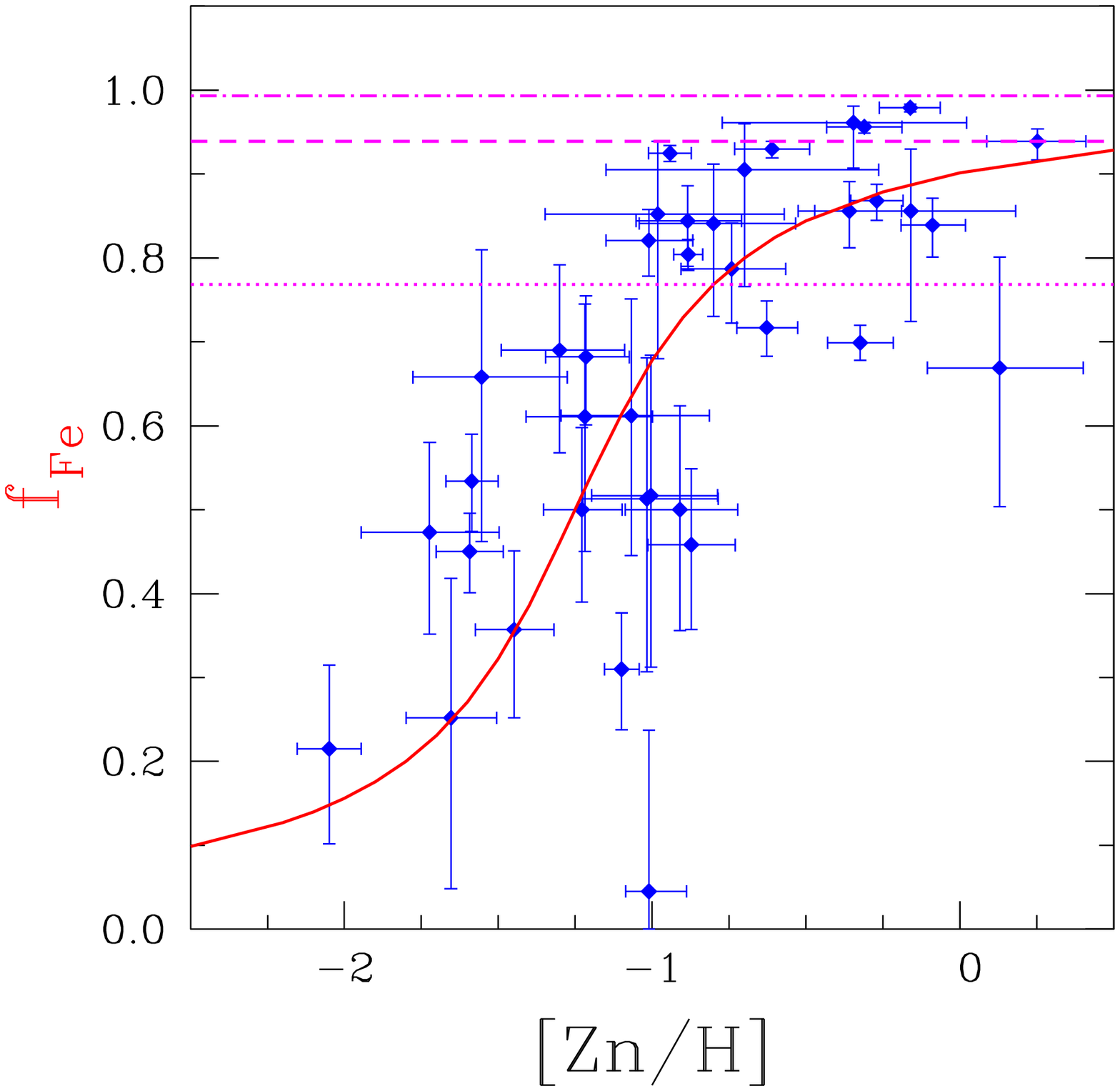}
  \caption{Left panel: V band extinction versus zinc column density
  in the interstellar medium (see Vladilo \& P\'eroux 2005).
  Right panel: Fraction of iron in dust versus metallicity in DLAs
  (Vladilo 2004); smooth line: analytical approximation adopted
  in the present study.
  }
  \label{AV_NZn}
\end{figure}

We tested and calibrated the relation (\ref{A_Zn}) using interstellar measurements
of extinction and $N$(ZnII).
The results are shown in Fig. \ref{AV_NZn}. 
The data follow the expected trend and yield a mean value
of the $a_\lambda$ parameter in V band $a_V = 0.29 (\pm 0.07) 10^{-13}$
magnitudes cm$^{-2}$. 
From the condition $\partial A_\lambda / \partial (\log N_\mathrm{Zn}) =1$ magnitude,
we obtain a turning point $\log N$(ZnII) $\simeq 13.18$, above which the
extinction experiences a dramatic rise. 
This extinction barrier lies remarkably close to the Boiss\'e threshold  
and this fact motivated us to study the extinction relation for DLAs. 
As a result, we obtained that the  
observer's frame extinction of a DLA at redshift $z$ is
\begin{equation}
A_{\lambda} (N_\mathrm{H},Z; z) 
\simeq A_\circ ~ G ~\, 
\xi \! \left(  \! { \lambda \over 1 \!+ \! z } \! \right)  ~ 
f_{\mathrm{Fe}}(Z) ~\, 
 N_{\mathrm{H}} ~ Z ~ , 
\label{AlamHzeta}
\end{equation}
where 
$A_\circ \simeq 1.85 \times 10^{-14}$ mag cm$^2$,
$G = (g_\lambda)_\mathrm{DLA}/(g_\lambda)_\mathrm{ISM}$,
and $\xi(\lambda)=A(\lambda)/A(\lambda_V)$
is a normalized extinction curve;
$Z =$ Zn/H and $N_{\mathrm{H}}$ are the metallicity and HI
column density of the absorber; 
$f_{\mathrm{Fe}}(Z)$ is an analytical function,
based on depletion data  (Vladilo 2004),
that gives the fraction of iron in dust as a function of the metallicity. 
Since $f_{\mathrm{Fe}}(Z)$ vanishes at low metallicities
(Fig. 1, right panel), the predicted extinction will accordingly vanish.
Thanks to this fact, we do not need to know  
 the value of the grain parameter at the very
early stages of chemical evolution, when the  behaviour of the dust
is most unpredictable. 
%
In the application of  Eq. (\ref{AlamHzeta}) we consider
both a `Milky-Way' and a  Small Magellanic Cloud (`SMC')  type of dust.
For each type we adopt an appropriate  
$\xi(\lambda)$ and       $G$,
making also use of SMC extinction data
(see details in the full paper).
 
\section{Method} 

At a given redshift, the   distribution of DLA  extinctions
and the obscuration bias
are completely determined from (\ref{AlamHzeta})
for any assigned pair of    $N$(HI) and $Z$ distributions.
%
%
%
%
We used this property to find the relation between
true and biased distributions of $N$(HI) and $Z$
 in DLAs.
Simple mathematical expressions can be derived by assuming that
the distributions of $N$(HI) and $Z$ are statistically independent and
by ignoring multiple DLA absorptions in a given line of sight. 
Both assumptions are conservative in the sense that they may
slightly underestimate the effect of quasar obscuration. 
 The fraction of DLA-quasar pairs that can be detected
in the differential element 
$d N_\mathrm{H} \, dZ\, dz$ 
is
\begin{equation}
B_{m_\ell}  ( N_\mathrm{H}, Z,z) \equiv
{
  \int_\circ^{m_{\ell}-A_\lambda(N_\mathrm{H},Z, z)} \,  
 n(m; z)  \, d m \,
   \over
  \int_\circ^{m_\ell}  \, n(m; z)  \,  dm 
 \label{BigBHZz}
 } \, ,
\end{equation}
where $A_\lambda(N_\mathrm{H},Z, z)$ is     
the extinction   (\ref{AlamHzeta}) ,
$n(m)$ is the   distribution of the quasar apparent magnitudes $m$,
and $m_\ell$ the limiting  magnitude of the survey.
In the redshift interval $z_1 \leq \overline{z} \leq z_2$ the relations between
the true   distributions  of $N$(HI) and $Z$ in DLAs, $f_{N_\mathrm{H}}$ and $f_{Z}$,
and  the corresponding biased distributios,
$f^\mathrm{b}_{N_\mathrm{H}}$ and
$f^\mathrm{b}_{Z}$, are
 \begin{equation}
 f^\mathrm{b}_{N_\mathrm{H}}  \simeq 
{
\int_{0}^{\infty}
  \mathcal{B}_{m_\ell}  ( N_\mathrm{H}, Z) \,
f_{Z}  \, d\, Z
 \over 
\int_{0}^{\infty}
f_{Z}  \, d\, Z
 } ~  f_{N_\mathrm{H}} ~~
 \label{fb_H}
\end{equation}
\begin{equation}
 f^\mathrm{b}_{Z}  \simeq   
{
\int_{N_\mathrm{DLA}}^{\infty} 
  \mathcal{B}_{m_\ell}  ( N_\mathrm{H}, Z) \,
f_{N_\mathrm{H}}  \, d\, N_\mathrm{H} 
 \over 
\int_{N_\mathrm{DLA}}^{\infty} 
f_{N_\mathrm{H} }  \, d\, N_\mathrm{H}
 } ~  f_{Z} ~~,
 \label{fb_zeta}
\end{equation}
where 
$\mathcal{B}_{m_\ell}  ( N_\mathrm{H}, Z) = 
B_{m_\ell}  ( N_\mathrm{H}, Z, \overline{z})$
and
$N_\mathrm{DLA}=10^{20.3}$ atoms cm$^{-2}$.
The goal of our procedure is to invert these equations in order
to recover the true distributions given the biased ones, which can be  
estimated from magnitude-limited surveys. To do so we use the
procedure sketched in Fig. 2. We model the
unknown   distributions $f_{N_\mathrm{H}}$ and $f_{Z}$
using   analytical functions with free   parameters.
We start from a set of trial parameters and compute the extinction
distribution,
the function $ \mathcal{B}_{m_\ell}  ( N_\mathrm{H}, Z)$ 
and the biased distributions. We then compute the
$\chi^2$ deviation between the   biased distributions and the
 observed ones. The procedure is iterated to find a minimum in the
 $\chi^2$ values. At each iteration we estimate   the true
 distribution of quasar apparent magnitudes, $n(m)$, by 
 correcting the observed magnitude distribution   taking
 into account the effect of the extinction.
In this way we do not need to assume a quasar luminosity
function. For this and other reasons our procedure is different from
the one presented by Fall \& Pei (1993).

\begin{figure}
 \hskip 1.5 cm
 \includegraphics[width=7.5cm,angle=-90]{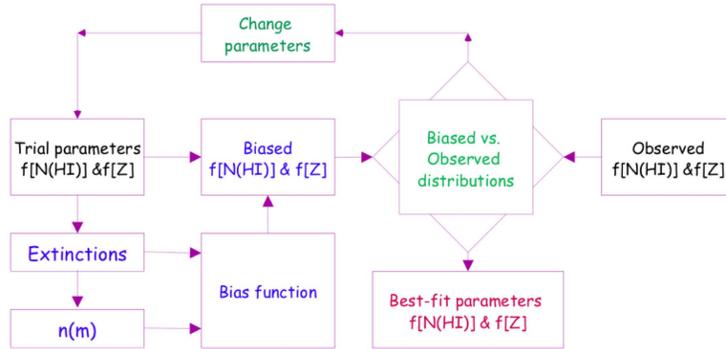}
 \vskip -1.2 cm
  \caption{ Flow diagram of the procedure adopted with the aim of
  recovering the true distributions of HI column densities
  and metallicities in DLAs starting from the  
  observed distributions. }
  \label{FlowDiagram}
\end{figure}

\section{Results}

As a first application of our method we considered the  
sample of DLAs with ZnII data. 
The   distribution of   apparent magnitudes was measured
from the quasar counts in the $r$ and $g$ bands of the Sloan Digital
Sky Survey. 
The extinctions were estimated in the same two bands.
The selected sample  
is characterized by  a quasar limiting magnitude $m_\ell \simeq 19$ in the optical,
required to detect the ZnII lines at  high spectra resolution.  
Most data are concentrated in the   range
$1.8 \leq z \leq 3.0$, with a median redshift $z  \approx 2.3$. 
The current limitation of the statistics (about 40 DLAs in 2004)
allowed us to compute the empirical distributions  
in a small number of bins.
%
Because of the limited empirical contraints
we modeled the true distributions $f_{N_\mathrm{H}}$ and $f_{Z}$
using functions
with the smallest possible number of   parameters.
For the $N$(HI)  distribution we adopted a   power law
$f_{N_\mathrm{H}} = C \times N_\mathrm{H}^{-\beta}$
truncated at the highest observed values of DLA column density, 
$\log N$(HI) $\simeq 22$. 
For the metallicity distribution we adopted a Schechter function
$f_{Z}= C^\prime \times (Z/Z_\ast)^\alpha e^{-Z/Z_\ast}$. 
%
With tighter observational
constraints available in the future, it will be possible to model  the true distributions
more realistically.  
%

%

For the different types of dust (MW- and SMC-type) and photometric bands
considered,
we found that the slope of the $N$(HI) power law lies in the range
$1.5 \lesssim \beta \lesssim   1.6$, with typical fit errors of $\pm 0.12$.
%
The mean value $< \!  Z \!>$ of the Schechter metallicity distribution lies in the range
$-0.4 \lesssim \log (<\!  Z \!> \! /Z_\odot) \lesssim -0.3$, with   fit errors of $\pm 0.15$ dex and total errors (including the uncertainty of the empirical distributions)
of $\pm 0.3$ dex.

The mean metallicity that we derive is higher 
than the canonical
value of DLAs taken at face value, $ <\! \mathrm{[Zn/H]} \! > \simeq -1.1$ dex.  
As we show in Fig. 3,
our estimate is consistent at $\approx 1 \sigma$ level
with the mean metallicity of the unbiased CORALS sample  (Ellison 2005).
In the full paper we show that also
our   predictions
of the    number density, $n_\mathrm{DLA}(z)$,
and   gas content, $\Omega_\mathrm{DLA}$, are  
consistent with CORALS results   (Ellison et al. 2001).  
%
%

\begin{figure}
 \hskip 0.7cm
  \includegraphics[width=5.5cm,angle=0]{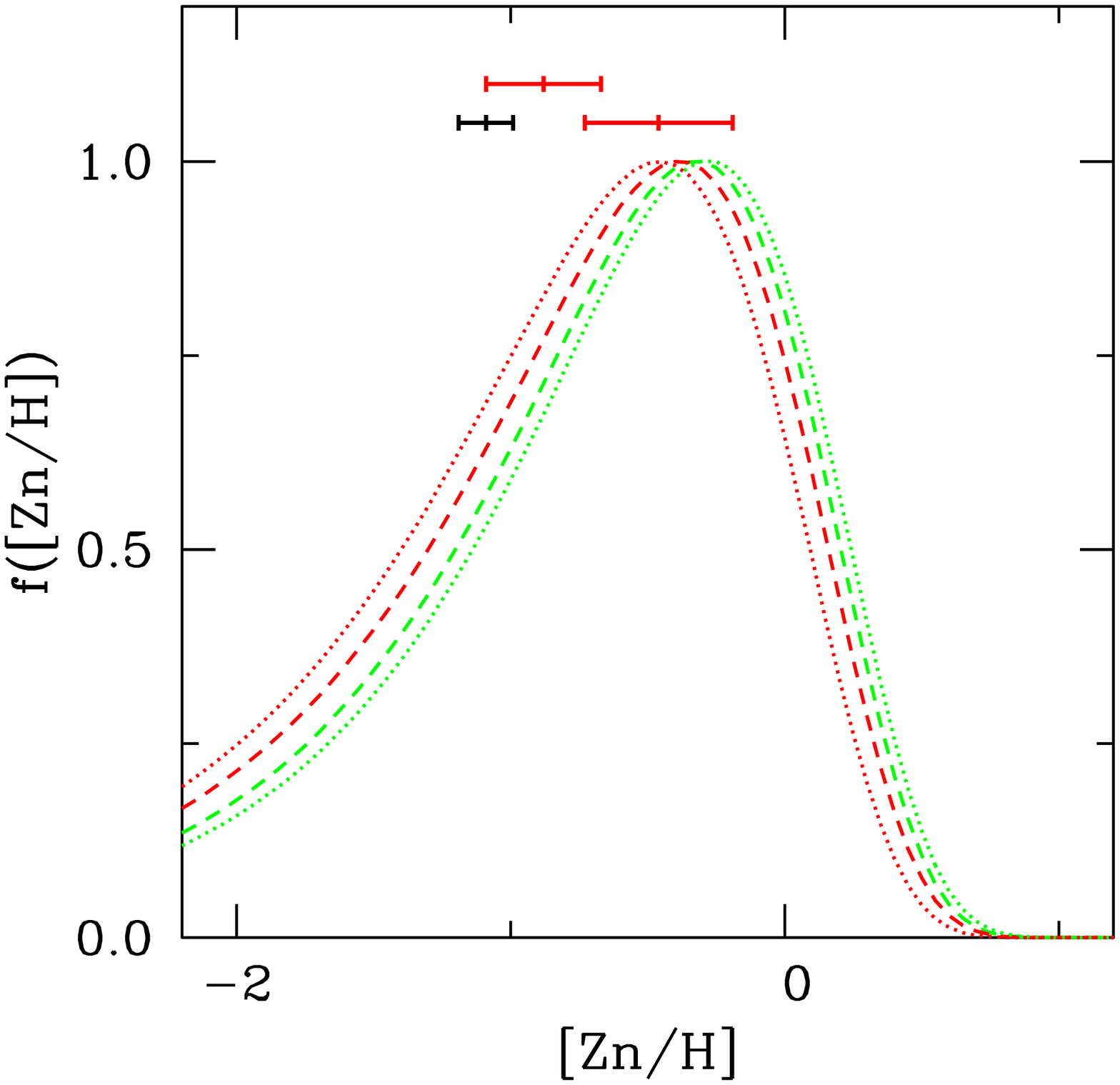}
 \hskip 0.7cm
 \includegraphics[width=5.2cm,angle=0]{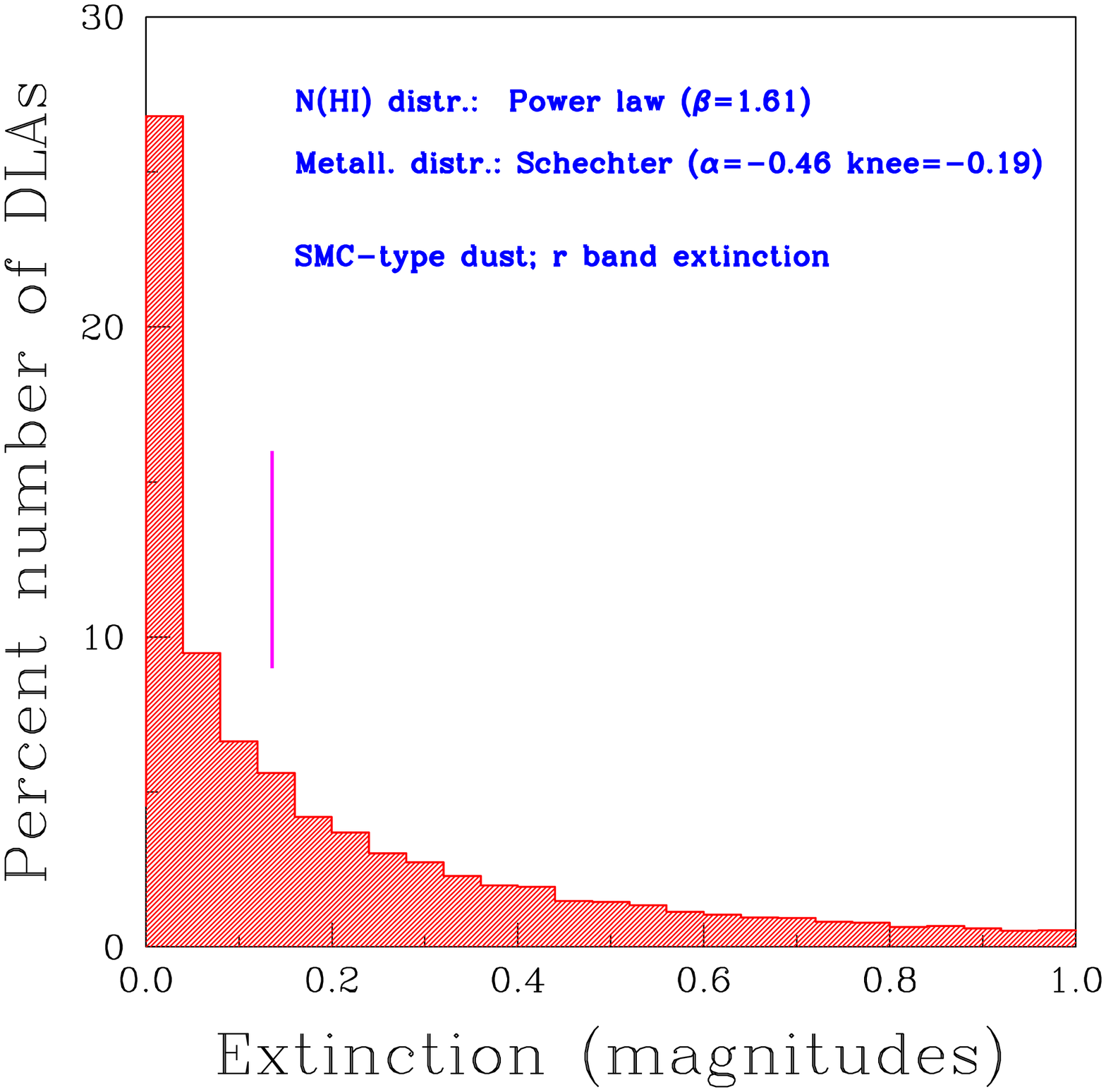}
  \caption{Left:   distribution of metallicities in DLAs
  for different predictions of our models
  (see Vladilo \& P\'eroux 2005); the error bars 
  indicate, from left to right, the $\pm 1 \sigma$ errors of the mean metallicity of DLAs
  estimated (1) at face value, (2) from the CORALS survey  (Ellison 2005),
  and (3) from our predictions. 
  Right:  distribution
  of DLA extinctions in the $r$ band obtained  
  for the ZnII sample of DLAs
  for an SMC-type dust; vertical bar: median extinction. }
  \label{ExtDistr}
\end{figure}

An example of distribution of DLA extinctions predicted by our study is
shown in    Fig. 3 (right panel).
The distribution has a median value of 0.14 mag in the
$r$ band, but is extremely asymmetric.
Most DLAs have negligible or little extinction, but the long tail
of the distribution, extending up to very high values, 
indicates that a significant fraction of DLAs
may indeed obscure the background quasar.
We estimate that the missed fraction is $\approx$ 30\% to 50\%
at limiting magnitude  $m_\ell =19$, but decreases steadily with $m_\ell$.
Surveys $m_\ell \gtrsim 21$ are weakly affected by the bias. 
%

The extinction distribution   highlights the difficulty of measuring
     the `mean' DLA reddening using large data sets of quasars.
%
The predicted median     reddening is
relatively low ($\simeq 0.05$ magnitudes) and
hard to detect on the uncertain quasar continuum.
In addition, the procedure of stacking a large
number of spectra for detecting the `mean' reddening
can give a meaningful answer only if the mean change of
continuum slope is a fair
representation of the change of slope of individual cases.
%

The metallicity dependence of the iron dust fraction  
(Fig. 1, right panel) indicates that DLA extinction should become
negligible at [Zn/H] $\lesssim -1.5$.
In conjunction with the trend of decreasing  [Zn/H]  with
redshift (Vladilo et al. 2000), this in turn suggests that DLA extinction
may vanish   at  $z \gtrsim 3$. 
Therefore, the difficulty of detecting quasar reddening from DLAs at $z \approx 2.8$  
(Murphy \& Liske 2004)  may   be due, in part, to the
intrinsic decline of the dust content at high redshift.

\begin{figure}
 \hskip 0.7cm
 \includegraphics[width=5.2cm,angle=0]{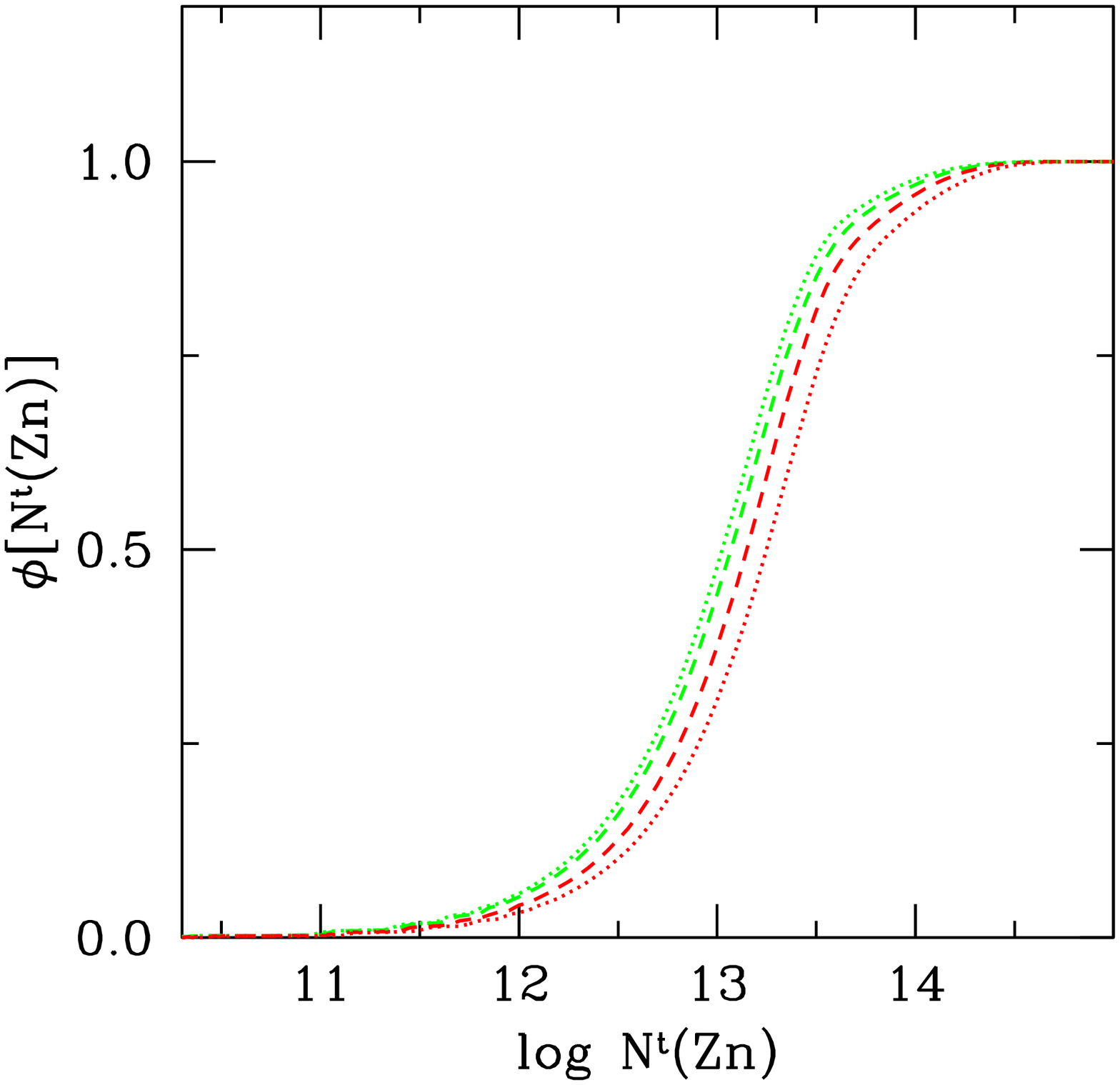}
 \hskip 0.7cm
 \includegraphics[width=5.2cm,angle=0]{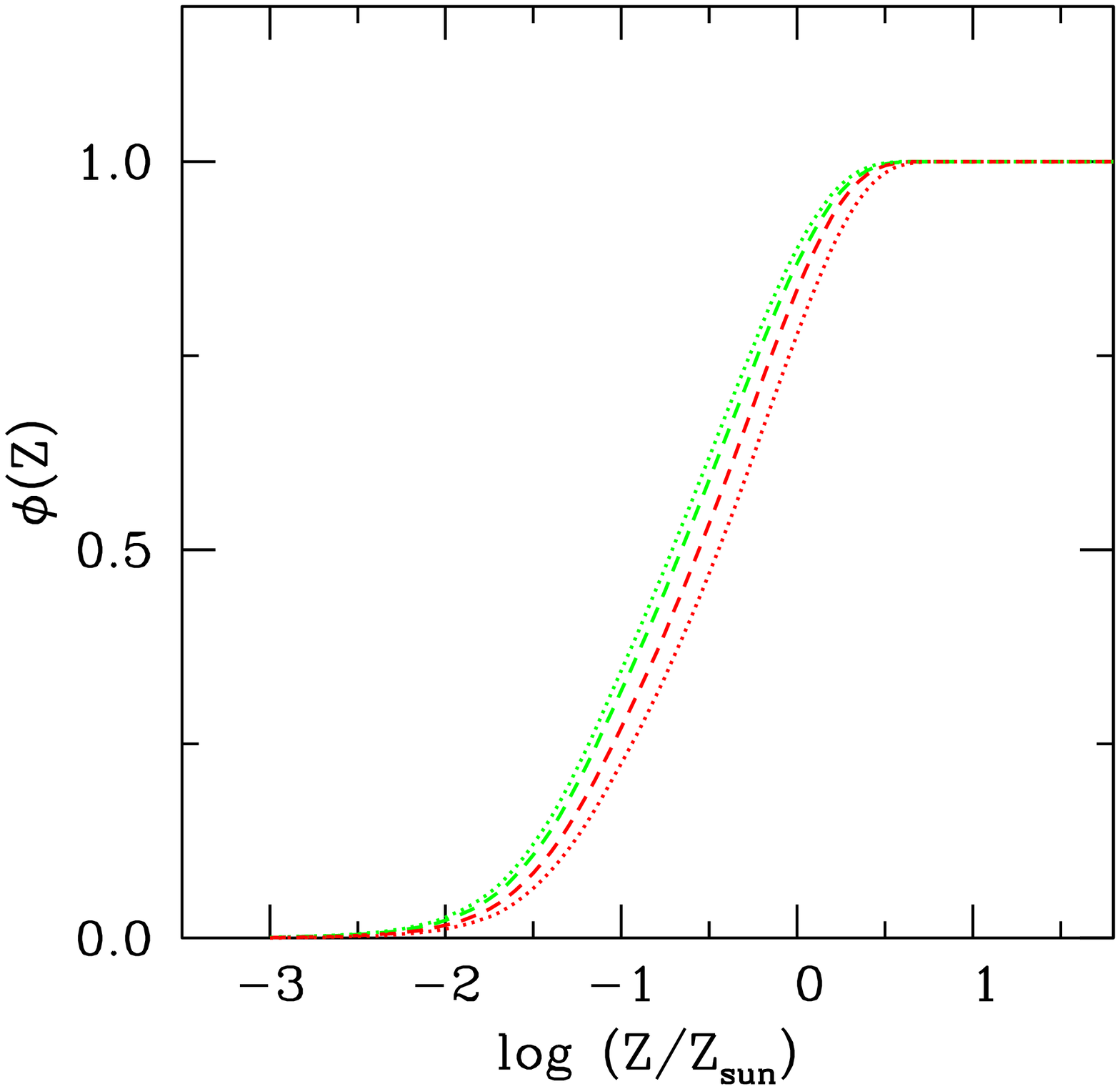}
  \caption{ Predicted fraction of obscured DLAs as a function
  of total Zn column density (left panel) and metallicity (right panel)
  in a DLA-quasar survey with limiting magnitude $\simeq 19$
  for different models (MW- and SMC-type dust;  
  $r$ and $g$ photometric bands).     }\label{phi}
\end{figure}

An interesting prediction of our method is the fraction
of obscured DLAs as a function of   $N$(ZnII), $\phi(N_\mathrm{Zn})$.
In   Fig. \ref{phi} (left panel) one can see that $\phi(N_\mathrm{Zn})$
shows a sharp rise exactly around the 
Boiss\'e's   threshold 
$\log N(\mathrm{ZnII}) \simeq 13.15$. 
This remarkable result, obtained
without  tuning   the dust parameters,   demonstrates 
that the obscuration effect provides 
a natural explanation for the    threshold. 
To discard this explanation one should 
 invoke some unknown mechanism for explaning
the lack of DLAs with $\log N(\mathrm{ZnII}) \gtrsim 13.15$
 and, in addition, one should tune the
dust parameters in such a way to move the threshold 
at much higher values of  $\log N(\mathrm{ZnII})$. 
Our explanation  is certainly more
attractive in terms of economy of hypothesis. 

The reality of Boiss\'e's  threshold has been invoked  
to bring into agreement predictions of galactic chemical evolution
models (Prantzos \& Boissier 2000) and of cosmological simulations
(Cen et al. 2003) with seemingly discrepant observations of DLAs. 
The existence of  a physical explanation for the  threshold
favours  a scenario in which the predictions of such models are generally
correct and the obscuration bias plays an important role. 

The combination of the   limit
$\log N(\mathrm{ZnII}) \leq 13.15$  with
the   cutoff 
$\log N(\mathrm{HI}) \geq 20.3$     imposed in the surveys  
creates a   cutoff of the highest possible metallicity
detectable in DLAs. 
The way the obscuration bias affects the metallicity distribution is
shown in the right panel of Fig. \ref{phi}. One can see that the bias
severely hampers the detection of DLAs with near solar metallicity. 
Because of the implications on our
understanding of the   star formation history of the Universe,
the   effect  on the metallicity distribution is probably
the most important consequence of the obscuration bias.

%
%
%
%


\begin{thebibliography}{}

\bibitem[1998]{B&98} 
{Boiss\'e, P., Le Brun, V., Bergeron, J., \& Deharveng, J.M.}
1998, \textit{A\&A} 333, 841

\bibitem[]{} 
{Cen, R., Ostriker, J.P., Prochaska, J.X., \& Wolfe, A.} 2003 \textit{ApJ}  598,  741

\bibitem[]{} 
{Ellison, S. L., Yan, L., Hook, I.M., Pettini, M., Wall, J.V., \& Shaver, P.}
2001, \textit{A\&A} 379, 393

\bibitem[]{}
{Ellison, S. L} 2005, \textit{These Proceedings}, p.\ xxx

\bibitem[]{} 
{Fall, S.M. \& Pei, Y.} 1993, \textit{ApJ} 402, 479 


\bibitem[]{}
{Khare, P., York, D.G., Vanden Berk, D., 
 et al.} 2005, \textit{These Proceedings}, p.\ xxx


\bibitem[]{}
{Lopez, S.} 2005, \textit{These Proceedings}, p.\ xxx

\bibitem[]{}
{Meurer, G.R.} 2004 in: A.N. Witt et al. (eds.),
\textit{Astrophysics of Dust, ASP Conf. Ser.} 309, 195

\bibitem[]{} 
{Murphy, M.T., \& Liske, J.} 2004, \textit{MNRAS} 354, L31  

\bibitem[]{} 
{Ostriker, J.P., \& Heisler, J.} 1984, \textit{ApJ} 278, 1

\bibitem[1991]{pfb91} 
{Pei, Y.C., Fall, S. M.,  \& Bechtold, J.} 1991, \textit{ApJ} 378, 6

\bibitem[]{} 
{ Prantzos, N. \& Boissier, S.} 2000, \textit{MNRAS} 315, 82

\bibitem[]{} 
{Spitzer, L.} 1978, 
\textit{Physical Processes in the Interstellar Medium}
(New York: Wiley Interscience)
 
 \bibitem[Vladilo (2004)]{Vlad04}
 {Vladilo, G.} 2004,  \textit{A\&A} 421, 479
 
 \bibitem[]{}
 {Vladilo, G., Bonifacio, P., Centuri\'on, M., \& Molaro, P.} 2000,  \textit{ApJ} 543, 24
 
 \bibitem[]{}
{Vladilo, G. \& P\'eroux, C.} 2005,  \textit{A\&A} in press,
(astro-ph/0502137)  
     

 
\end{thebibliography}
\end{document}